\documentclass[10pt,conference]{IEEEtran}
\IEEEoverridecommandlockouts
\usepackage{adjustbox}
\usepackage{cite}
\usepackage[most]{tcolorbox}
\usepackage{amsmath,amssymb,amsfonts}
\usepackage{booktabs}
\usepackage{tabularx}
\usepackage[table]{xcolor}
\usepackage{multirow}
\usepackage{url}
\usepackage{hyperref}
\definecolor{summaryBg}{RGB}{230,241,253}      
\definecolor{summaryBorder}{RGB}{83,142,204}  
\colorlet{quotelinecolor}{summaryBorder}

\newcommand{\quoteline}{%
  \hspace*{-\parindent}%
  \textcolor{quotelinecolor}{\vrule width 2pt}%
  \hspace{0.5em}%
}
\definecolor{qualityComp}{RGB}{244,244,244}
\definecolor{qualityEff}{RGB}{255,242,205}
\definecolor{qualityMaint}{RGB}{217,233,219}
\definecolor{qualityBarrier}{RGB}{218,232,250}
\usepackage{tabularx}
\usepackage{algorithmic}
\usepackage{booktabs}
\usepackage{graphicx}
\usepackage{textcomp}
\usepackage{xcolor}
\usepackage{tabularx}
\def\BibTeX{{\rm B\kern-.05em{\sc i\kern-.025em b}\kern-.08em
    T\kern-.1667em\lower.7ex\hbox{E}\kern-.125emX}}
\begin{document}



\title{Can AI Guide Testing Without Generating Tests? A Comparative Study of AI-Assisted Unit Testing}

\author{
\IEEEauthorblockN{
Zihan Fang,
Ron Asoulin-Handelman,
Helen Wu,
Yu Huang,
Gina R. Bai
}
\IEEEauthorblockA{
Vanderbilt University, Nashville, TN, USA\\
\{zihan.fang, ron.asoulin-handelman, helen.wu,
yu.huang, rui.bai\}@vanderbilt.edu
}
}


\maketitle

\begin{abstract}
Recent advances in generative AI have enabled assistants to generate increasingly sophisticated executable unit tests. However, novices may accept these outputs without independently identifying what to test, evaluating test-suite completeness, or verifying correctness, potentially limiting opportunities to develop essential testing skills. 
Although prior work has explored AI-assisted test generation and feedback-oriented programming tutors, little is known about whether AI can support novices' testing practices without generating tests for them.
To investigate this, we introduce TestCompass, a generative AI tutor that provides targeted feedback on weaknesses in students' test suites without generating executable tests. In a classroom study with 18 students, we compared TestCompass with a general AI assistant across two Java unit-testing tasks, examining students' testing behaviors, test-suite quality, and subjective experiences.
We found that participants using the general-purpose AI assistant produced more tests, whereas TestCompass participants spent more time developing and verifying tests themselves. 
However, most test-suite quality measures were comparable between conditions, and TestCompass participants achieved similar requirement coverage with significantly fewer test methods ($p=0.012$) in one task. Neither group reported short-term improvements in perceived testing knowledge or confidence. These findings suggest that feedback-oriented AI can support meaningful testing outcomes while preserving opportunities for hands-on testing practice, offering a complementary approach to AI-assisted testing education.
\end{abstract}


\section{Introduction}
\label{sec:intro}

Generative AI is reshaping both programming education and
real-world development workflows, changing how programmers
write code and the skills they need~\cite{denny2024prompt,tang2026programming}.
As AI-generated code becomes increasingly common, students
must also learn to critically evaluate and verify code
they did not write themselves~\cite{lau2023ban,lyu2026practitioners}.
Software testing plays a central role by providing a systematic way to determine whether generated code behaves as intended~\cite{barr2014oracle}, rather than accepting AI-generated solutions without adequately verifying their correctness~\cite{prather2024widening}. 
However, students can now also rely on generative AI to produce the tests used for verification, creating a new challenge: learners must evaluate not only AI-generated code but also the quality of AI-generated tests~\cite{yuan2024evaluating}. 
This is particularly concerning because software testing often receives limited instructional emphasis in computing curricula~\cite{astigarraga2010emerging}, and novices already struggle to decide what to test, determine whether a test suite is sufficient, and evaluate the correctness of their tests~\cite{shin2024model,bijlsma2021students,bai2021students}.

Prior work has explored how general-purpose AI assistants, such as ChatGPT, can support students in software testing~\cite{jalil2023chatgpt,ardic2026students,fang2025comparative}. 
These tools can explain testing concepts, suggest test cases, and generate executable tests, making testing more accessible~\cite{mohapatra2025artificial}. 
However, prior studies have raised concerns about incorrect or incomplete AI-generated tests, insufficient verification, and students' reliance on AI-generated solutions~\cite{jalil2023chatgpt,ardic2026students,yang2026large}. By providing ready-to-use tests alongside explanations, general-purpose AI assistants may encourage students to adopt generated solutions without critically evaluating their correctness or reasoning through testing decisions themselves~\cite{ardic2026students}.
This motivates exploring AI as a tutor that guides students' testing decisions rather than providing ready-to-use solutions. An important question is whether such an approach can provide meaningful testing support while keeping students actively engaged in designing and verifying their own test suites.

Prior work has explored feedback-oriented AI tutors for programming that provide hints and explanations without revealing solutions~\cite{liffiton2023codehelp,bassner2024iris}, as well as conceptual feedback and structured scaffolding for software testing~\cite{cordova2021comparison,yang2026large,perretta2025instructor}. Other studies have compared general-purpose AI assistants with testing checklists or pedagogically designed tutors~\cite{fang2025comparative,bassner2025contextaware}. However, evidence remains limited on how testing-specific, feedback-only generative AI compares with general-purpose AI assistants that generate executable tests, particularly in terms of novices' testing behaviors and test-suite quality. Understanding this distinction can inform the design of AI tools that support effective testing while keeping students actively engaged in test design and verification.

We investigate this gap in the context of unit testing, a foundational practice in programming and software engineering education~\cite{desai2009tdd,heckman2020testing} that requires students to understand expected behavior, identify edge cases, and assess whether their tests adequately cover intended functionality~\cite{bijlsma2021testing}. 
We introduce TestCompass, a testing-specific generative AI tutor that provides targeted feedback on weaknesses in students' test suites without generating tests for them. 
In a classroom study with 18 students, we compare TestCompass with a general AI across two Java unit-testing tasks, examining students' testing behaviors, test-suite quality, and experiences with the tools. 
General AI participants spent more time interacting with AI assistants, whereas TestCompass participants devoted more time to developing and verifying tests themselves based on AI-generated feedback. Most test-suite quality measures were comparable between conditions, while TestCompass participants achieved similar requirement coverage with significantly fewer test methods. 
We make the following contributions:
\begin{itemize}
    \item TestCompass, a feedback-oriented generative AI
    tutor that guides novice unit testing without
    generating executable tests.
    \item An empirical comparison of feedback-only and general-purpose AI assistants in terms of students' testing behaviors, test-suite quality, and perceptions.
    \item Design insights for future AI tutors in software testing education.
\end{itemize}

Overall, TestCompass demonstrates the potential of AI to support effective testing without taking test development away from students, while highlighting the need for future AI tutors to provide greater guidance on test implementation and verification.

\section{Related Work}
\label{sec:related-work}

\subsection{LLM-Based Support for Software Testing}
\label{sec:related-work-1}

Large language models (LLMs) have increasingly been
used to automate software testing, particularly
the generation of unit tests. Prior studies have
evaluated the effectiveness of different LLMs
and compared them with traditional automated
testing techniques, including search-based
software testing and symbolic execution
~\cite{abdullin2025testwars,
yang2024evaluationLLMsunittests,
konstantinou2026howwell}.
These studies demonstrate the potential of
LLMs for generating executable tests while
identifying limitations in coverage and correctness.

To improve test generation, researchers have
proposed frameworks that combine LLMs with
additional program context, iterative refinement,
and automated validation
~\cite{chen2024chatunitest,lemieux2023codamosa,
zhang2025large,kanagaraj2025llmdriven}.
For example, CodaMOSA~\cite{lemieux2023codamosa}
combines LLM-generated tests with search-based
testing, while ChatUniTest~\cite{chen2024chatunitest}
uses contextual information and iterative
validation to improve generated tests.
Yuan et al.~\cite{yuan2024evaluating}
systematically evaluated ChatGPT-generated
unit tests and identified correctness issues,
including compilation errors and incorrect
assertions. Their findings highlight the
importance of validating AI-generated tests
rather than relying solely on their apparent
plausibility.
While these approaches primarily aim to
improve the quality of automatically generated
tests, our work examines a different role
for generative AI: providing targeted
feedback that helps students develop and
verify their own test suites.

\subsection{Learning and Teaching Software Testing}
\label{sec:related-work-2}

Learning to write effective unit tests remains
challenging for novice programmers.
Prior research has identified difficulties
in selecting appropriate test cases,
understanding coverage criteria, and
determining whether tests adequately capture
program requirements
~\cite{zhang2024whystudentswrite,
andleeb2025enhancing,bai2021howstudents}.
These challenges highlight the need for
instructional support that helps students
reason about test design and evaluate the
quality of their test suites.
Researchers have explored different forms of
feedback to support students in developing
better tests. Cordova et al.
~\cite{cordova2021comparison} compared
inquiry-based conceptual feedback with
traditional detailed coverage feedback.
They found that conceptual feedback was
associated with higher coverage and fewer
redundant tests. More recently, Perretta
et al.~\cite{perretta2025instructor}
developed an automated feedback system that
provides instructor-written hints about
mutants not detected by students' tests.
Their classroom evaluation found a small
but statistically significant improvement
in students' ability to detect mutants.
These studies demonstrate the potential
of targeted feedback to guide test
improvement without providing complete
test implementations.

With the growing availability of generative
AI, recent studies have also investigated
how students use general-purpose AI
assistants for software testing.
Jalil et al.~\cite{jalil2023chatgpt}
examined ChatGPT's ability to answer
software testing questions, identifying
both opportunities and limitations for
testing education.
Fang et al.~\cite{fang2025comparative}
compared ChatGPT, testing checklists, and
their combination, finding comparable
testing performance across conditions
while raising concerns about students'
limited verification of AI-generated
responses. Ardic et al.
~\cite{ardic2026students} identified
different strategies through which
students collaborate with generative AI
in test ideation and implementation,
highlighting concerns about trust,
test quality, and ownership.
These studies demonstrate
the importance of both feedback and
students' engagement in testing.
However, they provide limited evidence
on how dynamically generated,
testing-specific feedback compares
with general-purpose AI assistance
that can directly generate executable
tests.

\subsection{Pedagogical AI Support in Programming Education}
\label{sec:related-work-3}

Generative AI has increasingly been
integrated into programming education
to provide students with accessible,
personalized assistance.
Prior studies have examined how novices
interact with general-purpose AI coding
assistants~\cite{prather2023itsweird},
while educational deployments such as
the AI-powered tools introduced in CS50
~\cite{liu2024cs50withAI}
have explored course-specific assistance
at scale.
To address concerns about students'
reliance on AI-generated solutions,
researchers have developed pedagogically
designed AI tutors that emphasize guidance
rather than direct code generation.
CodeHelp~\cite{liffiton2023codehelp}
uses prompting strategies and guardrails
to provide explanations and programming
guidance without revealing complete
solutions. Similarly, CodeAid
~\cite{kazemitabaar2024codeaid}
provides conceptual explanations,
pseudocode, and suggestions for
improving students' code while
restricting direct solutions.
Iris~\cite{bassner2024iris}
further demonstrates how context-aware
AI tutors can provide assistance
tailored to programming coursework.
Recent research has also compared
pedagogically designed AI tutors with
general-purpose AI assistants.
Bassner et al.
~\cite{bassner2025contextaware}
compared a context-aware AI tutor,
ChatGPT, and no AI assistance.
They found no significant differences in learning
gains or task performance, but observed differences
in students' preferences and verification behaviors.
Building on this work, we investigate how
feedback-only generative AI compares with
general-purpose AI assistance in unit testing,
focusing on students' testing behaviors,
test-suite quality, and experiences.

\section{Methodology}

\subsection{TestCompass Design}
\label{sec:testcompass-design}

\begin{figure}[t]
    \centering
    \includegraphics[width=0.75\columnwidth]{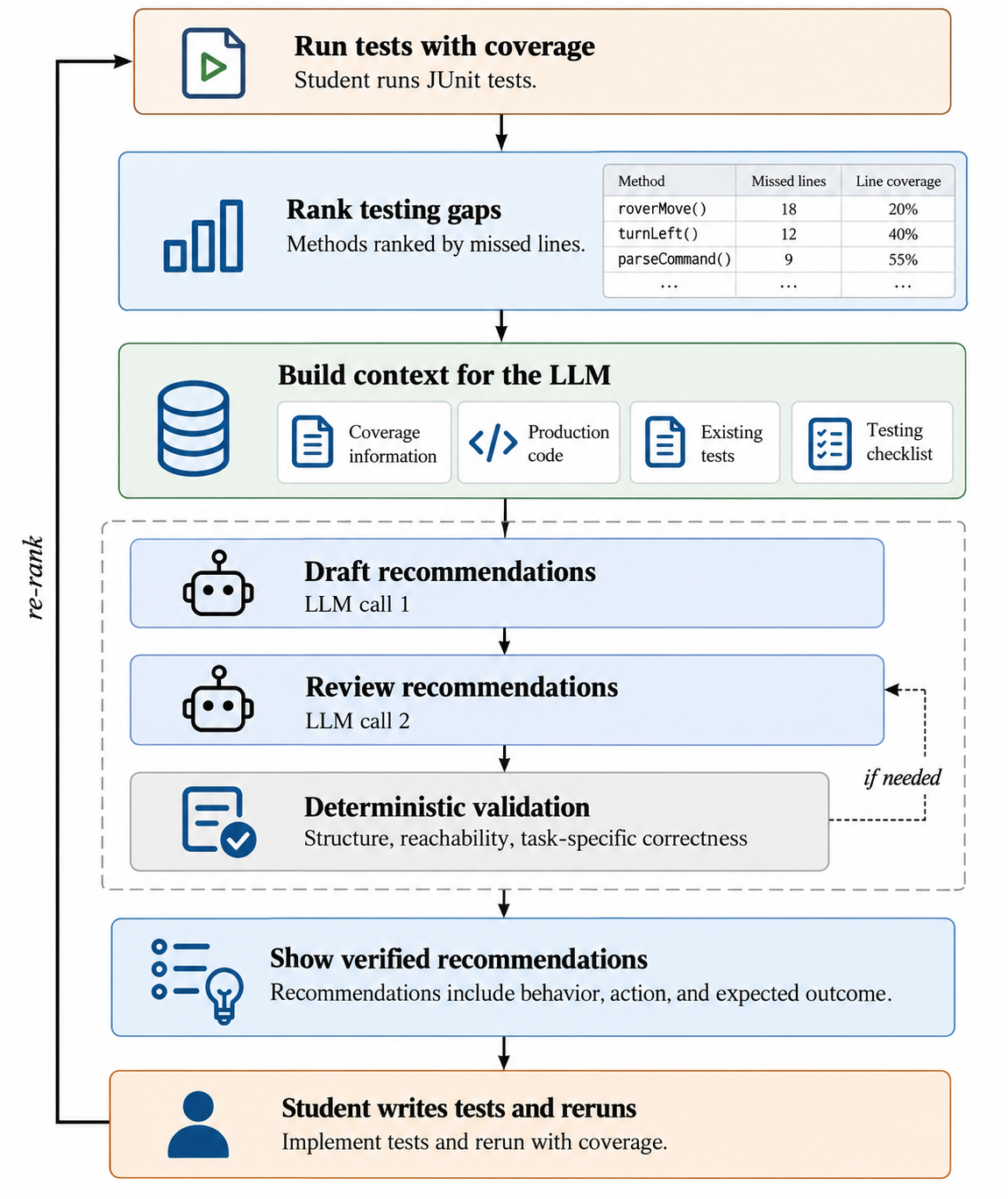}
    \caption{TestCompass Design}
    \label{fig:testcompass}
\end{figure}

TestCompass is an IntelliJ IDEA plugin designed to
help students identify what to test without generating test code for them (Fig.~\ref{fig:testcompass}).
The tool combines coverage analysis with generative AI:
coverage data identifies untested portions of the code,
while the LLM uses this information to generate targeted,
actionable feedback. Following prior work on conceptual
feedback in testing education~\cite{cordova2021comparison},
TestCompass aims to guide students' testing decisions
while keeping them actively involved in writing and
verifying their own tests.

After students run their tests with coverage analysis,
TestCompass ranks production methods by the number of
uncovered lines, using lower line coverage to break ties.
When students request feedback, the tool selects all ranked methods with uncovered lines.
For these methods, TestCompass collects coverage
information, relevant production code, and existing
test files to provide the LLM with context about
the program's behavior and what students have
already tested.
The prompt also includes a structured testing checklist
that instructs the LLM to identify meaningful untested
behavior, focus each recommendation on a single behavior,
specify an action and its expected outcome, and avoid
duplicating existing tests. 
This checklist guides the
feedback generation process but is not shown directly
to students. Based on this context, TestCompass generates
recommendations describing three elements: the behavior
to test, the action needed to exercise that behavior,
and the expected observable outcome. The prompt explicitly
prohibits generating executable test code. Students
then translate the recommendations into JUnit tests,
implement and execute them, and rerun coverage analysis.
The updated coverage results allow TestCompass to
identify remaining testing gaps, supporting an iterative
feedback process.

Before presenting recommendations to students, TestCompass
applies a multi-stage validation process. An initial
LLM call generates the recommendations, followed by
a second LLM pass that reviews them against the
production code, existing tests, and coverage context.
The reviewed recommendations then undergo deterministic
checks for the required feedback structure and,
for the study tasks, task-specific properties.
Recommendations that fail these checks trigger a
correction pass before being displayed. This process
aims to reduce incorrect or internally inconsistent
feedback and improve the reliability of the
generated recommendations.

\subsection{TestCompass Feedback Evaluation}
\label{sec:feedback-evaluation}
To assess TestCompass’s feedback, we developed a rubric informed by prior work on agent feedback evaluation~\cite{fang2025dpo}, conceptual feedback in testing education~\cite{cordova2021comparison}, and developer expectations of AI coding assistants~\cite{lyu2025my}. It covers five dimensions: (1) \textit{Correctness}, whether recommendations accurately describe program behavior and expected outcomes; (2) \textit{Relevance}, whether they address meaningful testing gaps without duplicating existing tests; (3) \textit{Actionability}, whether they give enough guidance to construct appropriate tests; (4) \textit{Clarity}, whether they are unambiguous for novices; and (5) \textit{Pedagogical Usefulness}, whether they encourage independent testing decisions rather than provide ready-made solutions.

We applied the rubric through expert review and three pilot sessions. 
Each dimension was rated on a five-point scale. Two researchers independently checked the recommendations against the task specifications, production code, existing tests, and coverage data, resolving disagreements through discussion. Three pilot sessions with participants experienced in software testing and computer science teaching followed. TestCompass averaged above 4 on every dimension across both evaluations.

\subsection{Study Tasks}
\label{sec:tasks}
\subsubsection{Unit-Testing Tasks}
We adapted two Java unit-testing tasks from a prior
study~\cite{fang2025comparative}, modifying the first
task from written test design to executable JUnit
test implementation. 
The tasks emphasized complementary
forms of testing reasoning while keeping the programming language, testing framework, and development environment
constant. For both tasks, participants received a complete implementation, behavioral specifications, and minimal starter JUnit tests. They were instructed
to modify only the designated test files and test
the provided code as thoroughly as possible.
All participants completed the tasks in a fixed order,
with recommended time limits of 40 minutes for
\textit{MarsRover} and 60 minutes for
\textit{BowlingScoreKeeper}.

\paragraph{MarsRover}
This first task involved testing a stateful \texttt{MarsRover}
API with 12 requirements covering rotation, movement,
command sequences, boundary wrapping, obstacle handling,
output formatting, and invalid inputs.
Participants extended \texttt{MarsRoverTest.java}
to test state transitions, boundary conditions,
and interactions among commands.

\paragraph{BowlingScoreKeeper}
The second task involved testing a bowling scoring system
implemented through the \texttt{Frame} and
\texttt{BowlingGame} classes.
Its 13 requirements covered frame scoring, strikes,
spares, bonus calculations, the ten-frame limit,
and special rules for the final frame.
Participants extended \texttt{FrameTest.java}
and \texttt{BowlingGameTest.java}, focusing on
dependencies across frames, where later throws
affect earlier frame scores.

\subsubsection{Surveys}
Participants completed a pre-study survey assessing
their self-reported Java unit-testing experience and
confidence in writing comprehensive tests. 
The post-study
survey collected demographic information, prior programming
and testing experience, perceived testing challenges,
tool-use behaviors, self-rated testing knowledge and
confidence, tool helpfulness, willingness to recommend
the tool, and open-ended feedback on its limitations
and potential improvements. 

\begin{figure*}[t]
    \centering
    \includegraphics[width=0.75\textwidth]{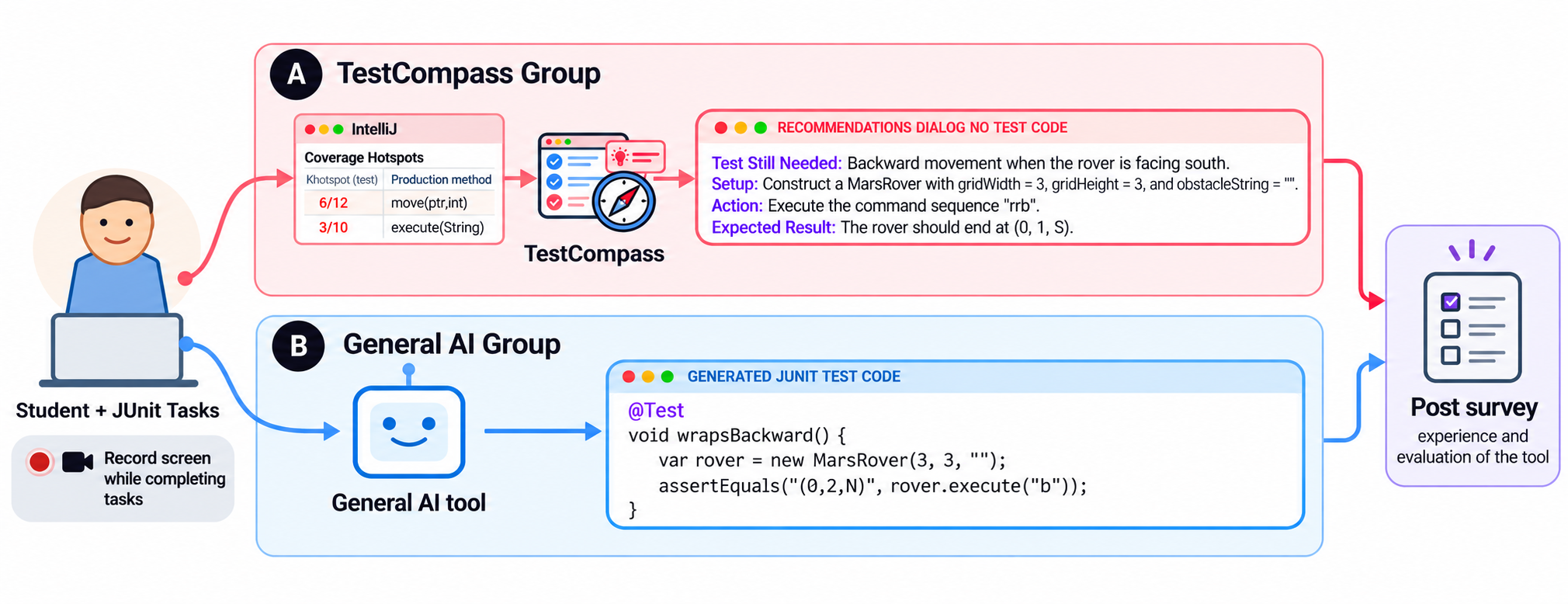}
    \caption{Study Design}
    \label{fig:study_design}
\end{figure*}

\subsection{Participant Recruitment and Group Assignment}
\label{sec:participants}

We initially recruited participants from a sophomore-level
computer science course at our institution and later
expanded recruitment to a senior-level course due to
limited participation. Participation was voluntary,
with four extra-credit points awarded toward a course
assignment upon study completion.
Eligible participants were required to be at least 18
years old, enrolled in one of the courses, have prior
Java programming experience, and have received
introductory instruction in unit testing and JUnit.
The study was conducted outside regular class
activities, and non-participation carried no penalty.

A total of 18 students participated and were randomly
assigned equally to TestCompass (TC) or general-purpose
AI (AI). However, one participant assigned to TC used
only general-purpose AI and was analyzed according to
the tool used, resulting in eight TC and ten AI
participants. No statistically significant differences
were observed between the analysis groups in education,
gender, programming experience, or Java experience.
All participants reported less than one year of
unit-testing experience. Most self-identified as
novices, while two identified as advanced beginners.
Table~\ref{tab:participant_demographics}
summarizes participant demographics and prior experience.

\begin{table}[t]
\centering
\caption{Participant demographics and self-reported
experience by condition.}
\label{tab:participant_demographics}

\footnotesize
\setlength{\tabcolsep}{2pt}
\renewcommand{\arraystretch}{1.15}

\resizebox{0.85\columnwidth}{!}{%
\begin{tabular}{@{}l*{9}{c}@{}}
\toprule
& \multicolumn{2}{c}{\textbf{Education}}
& \multicolumn{4}{c}{\textbf{Gender}}
& \multicolumn{2}{c}{\textbf{Java (yrs)}}
& \multicolumn{1}{c}{\textbf{Unit test}} \\
\cmidrule(lr){2-3}
\cmidrule(lr){4-7}
\cmidrule(lr){8-9}
\cmidrule(l){10-10}
\textbf{Groups}
& \textit{UG}
& \textit{Grad}
& \textit{M}
& \textit{W}
& \textit{NB}
& \textit{ND}
& \textit{$<1$}
& \textit{1--3}
& \textit{$<1$ yr} \\
\midrule
TestCompass (TC)
& 7 & 1 & 6 & 1 & 0 & 1 & 4 & 4 & 8 \\
General AI (AI)
& 10 & 0 & 5 & 4 & 1 & 0 & 4 & 6 & 10 \\
\bottomrule
\end{tabular}%
}

\vspace{0.3em}
\begin{minipage}{\columnwidth}
\scriptsize
\raggedright
\textit{Note.} Values represent participant counts.
UG = undergraduate; Grad = graduate;
M = man; W = woman; NB = non-binary;
ND = prefer not to disclose.
\end{minipage}

\end{table}

\subsection{Study Protocol}
\label{sec:study-protocol}

Our study was reviewed and approved by our institution's
Institutional Review Board (IRB). Before beginning,
participants provided informed consent and completed
a pre-study survey through Google Forms. They then
received a condition-specific GitHub repository
containing the project template, requirements
specification, and instructions for using their
assigned AI tool.
Participants in the TC condition were
instructed to use only TestCompass for AI assistance,
whereas those in the AI condition
could use a general AI assistant (e.g., ChatGPT) of their choice. 
Participants could install TestCompass directly from the JetBrains Marketplace or manually using a local plugin file.
We recorded the specific assistant used by each participant in the AI condition.
Participants completed both tasks in IntelliJ IDEA,
where they inspected the provided source code and
starter tests and developed JUnit tests based on
the requirements. They could iteratively execute
tests, examine failures and coverage results,
revise their test suites, and consult permitted
online resources. 
Screen recording was required throughout both tasks
to capture testing activities and interactions with
AI tools. After each task, participants submitted
their completed project files and screen recordings.
Upon completing both tasks, they completed a
post-study survey.
The study procedure is illustrated in
Fig.~\ref{fig:study_design}. Each study session lasted approximately 1.5 hours.

\subsection{Data Analysis}
\label{sec:data-analysis}
We analyzed participants' screen recordings, submitted test suites, and survey responses to examine how they interacted with different forms of AI support, their testing performance, and their perceptions of the tools. The dataset included 18 participants (eight TC and ten AI), 36 test suites containing 798 identified test methods, 36 screen recordings, and matched pre- and post-study survey responses from all participants.

\subsubsection{Interaction Analysis}
We analyzed 22.7 hours of screen recordings,
dividing them into event-based segments representing
observable activities. Idle, off-task, and unclassifiable
intervals were excluded. Activities included reading
task instructions and requirements, inspecting source
code, writing or revising tests, running or inspecting
JUnit results, consulting TestCompass or general-purpose
AI, using reference materials, performing web searches,
and setup-related activities (Table~\ref{tab:interaction}).
One researcher annotated all recordings with assistance
from GPT-6 to extract on-screen text, timestamps, and
supporting evidence. The researcher manually determined
segment boundaries and activity labels, verifying
the AI-assisted information against the original
recordings. 
A second researcher independently reviewed
annotations for half of the participants against the
recordings. Disagreements were resolved through
discussion until consensus was reached.
For each participant and task, we calculated the
frequency and cumulative duration of each activity.
To account for differences in task duration, we
calculated each activity's proportion of classified
task time at the participant level and then averaged
these proportions within each condition.

\subsubsection{Test-Suite Quality Analysis}

Following prior work on unit-testing evaluation
~\cite{fang2025comparative}, we evaluated each
submitted test suite across four dimensions:
\textit{completeness}, \textit{effectiveness},
\textit{maintainability}, and
\textit{implementation barriers}.
These dimensions capture how thoroughly tests
exercise intended behavior, detect faults,
maintain code quality, and reflect difficulties
in implementing correct tests.

\paragraph{Completeness}
We assessed completeness using \textbf{requirement,
line, and branch coverage}. Requirement coverage
was manually coded based on whether at least
one test addressed each specified requirement,
regardless of implementation errors or incorrect
expected results. Coverage was assessed against
12 requirements for \textit{MarsRover} and
13 for \textit{BowlingScoreKeeper}.
Line and branch coverage were measured using
the IntelliJ IDEA coverage agent. We also
repeated these measurements using JaCoCo
as a sensitivity analysis to assess whether between-condition comparisons depended on the coverage engine.

\paragraph{Effectiveness}
We assessed fault-detection effectiveness using \textbf{mutation
coverage} from PIT (PITest), defined as the proportion of
generated mutants killed by the submitted test suite.

\paragraph{Maintainability}
We evaluated maintainability using \textbf{SonarQube
static-analysis issues} and \textbf{test-name quality}.
We recorded both the total number of issues
and issues per test method to account for
differences in test-suite size. Two researchers
manually rated test names on a five-point
scale based on clarity, descriptiveness,
and meaningfulness, with ratings averaged
at the participant level. 
We also examined \textbf{redundancy} within each participant's
submitted test suite, identifying extra exact duplicates
and test methods whose setup, actions, and assertions
were fully covered by another method.

\paragraph{Implementation Barriers}
We examined difficulties in implementing executable tests
by recording the number of submitted test methods and
identifying erroneous tests. A test was classified as
erroneous if it failed to compile, did not match its
apparent testing intention, contained an incorrect
assertion or other logical error, or mishandled
exceptions. 
We report the \textbf{number of inventoried test methods} and the \textbf{percentage of erroneous tests} among all submitted test methods.

\subsubsection{Survey Analysis}
\label{sec:survey}
Closed-ended survey responses were summarized using percentages, means, and
standard deviations. Open-ended responses were analyzed using inductive
thematic analysis~\cite{braun2012thematic}. One researcher reviewed all responses,
generated initial codes, and grouped related codes into recurring
themes. 
Themes were iteratively refined until they captured recurring patterns
across responses. A second researcher reviewed the codes and resulting themes
for consistency, and disagreements were resolved through discussion. 
All 18 participants completed matched pre- and
post-study surveys.
\subsubsection{Statistical Analysis}
\label{sec:statistical}
Between-condition comparisons were conducted at the
participant level, separately for each task where
applicable. Given the small sample size, continuous
outcomes were compared using two-sided exact permutation
tests based on the Mann--Whitney $U$ statistic,
with Cliff's delta reported as the effect size.
Categorical outcomes were compared using Fisher's
exact test. We used a significance threshold of
$\alpha=0.05$ and applied Holm correction for
multiple comparisons.

\begin{table}[t]
\centering
\caption{Time allocation across testing activities
by task and tool condition.}
\label{tab:interaction}

\footnotesize
\setlength{\tabcolsep}{4pt}
\renewcommand{\arraystretch}{1.0}

\resizebox{0.85\columnwidth}{!}{%
\begin{tabularx}{1.0\columnwidth}
{@{}>{\raggedright\arraybackslash}X|rr|rr@{}}
\hline
\multirow{3}{*}{\textbf{Activity}}
& \multicolumn{4}{c}{\textbf{Average Time and Percentage}} \\
\cline{2-5}
& \multicolumn{2}{c|}{\textbf{TC}}
& \multicolumn{2}{c}{\textbf{AI}} \\
\cline{2-5}
& \textit{Min.} & \textit{\%}
& \textit{Min.} & \textit{\%} \\
\hline\hline

\multicolumn{5}{c}{\textit{Task 1: MarsRover}} \\
\hline

Tool consultation$^{*}$
& \textbf{2.3} & 9.2
& \textbf{13.5} & 41.2 \\

Web search
& 0.4 & 1.3 & 0.1 & 0.3 \\

References
& 0.3 & 0.7 & 0.3 & 1.0 \\

Requirements$^{*}$
& \textbf{0.2} & 0.5
& \textbf{3.0} & 9.4 \\

Source inspection
& 1.3 & 3.1 & 1.8 & 4.6 \\

Test work
& 22.3 & 72.6 & 11.1 & 31.8 \\

JUnit run/inspection
& 2.9 & 9.2 & 3.7 & 9.5 \\

Setup/other
& 1.0 & 3.4 & 0.8 & 2.2 \\

\textit{Total classified}
& 30.7 & 100.0
& 34.3 & 100.0 \\

\hline

\multicolumn{5}{c}
{\textit{Task 2: BowlingScoreKeeper}} \\
\hline

Tool consultation$^{*}$
& \textbf{2.5} & 8.2
& \textbf{12.4} & 52.6 \\

Web search
& 0.9 & 1.8 & 0.6 & 1.6 \\

References
& 2.0 & 7.5 & 0.7 & 2.0 \\

Requirements
& 1.7 & 4.7 & 1.6 & 5.4 \\

Source inspection
& 2.7 & 5.6 & 0.8 & 2.7 \\

Test work$^{*}$
& \textbf{27.1} & 59.1
& \textbf{8.9} & 22.9 \\

JUnit run/inspection
& 4.3 & 10.0 & 2.8 & 7.7 \\

Setup/other
& 1.0 & 3.1 & 2.4 & 5.1 \\

\textit{Total classified}
& 42.2 & 100.0
& 30.2 & 100.0 \\

\hline
\end{tabularx}%
}

\vspace{0.4em}

\begin{minipage}{\columnwidth}
\scriptsize
\raggedright
\textit{Note.}
Values are mean minutes and mean participant-level percentages within each condition.
Test work includes writing, revising, and
inspecting tests.
$^{*}$ Indicates a significant between-condition difference
(Holm-adjusted $p<0.05$).
\end{minipage}

\end{table}

\section{Results}
\label{sec:results}
We examined how participants engaged with different forms of AI support by analyzing their testing behaviors, test-suite quality, and self-reported experiences.
\subsection{RQ1: How Do Students Interact with Different AI Tools for Unit Testing?}
\label{sec:rq1}

\paragraph{Distribution of Observed Activities}
Participants in the two conditions showed distinct patterns of time allocation between hands-on test development and AI tool interaction.

Across both tasks, TC participants devoted a greater proportion of their time to test-related activities. In MarsRover, test-related work accounted for 72.6\% of their time, compared with 31.8\% in the AI condition. A similar pattern emerged in BowlingScoreKeeper (59.1\% vs. 22.9\%), where TC participants also spent significantly more time working on tests (27.1 vs. 8.9 minutes; $p=0.025$).
In contrast, AI participants spent substantially more time consulting their assistants. Tool consultation accounted for 41.2\% and 52.6\% of their time in MarsRover and BowlingScoreKeeper, respectively, compared with 9.2\% and 8.2\% in TC. The differences in consultation time were statistically significant for both MarsRover (13.5 vs. 2.3 minutes; $p=0.015$) and BowlingScoreKeeper (12.4 vs. 2.5 minutes; $p=0.011$).

These findings highlight how different forms of AI support may shape students' engagement in the testing process, particularly the balance between practicing test development and relying on AI assistance. Table~\ref{tab:interaction} summarizes the distribution of observed activities.

\paragraph{Patterns of AI Assistance}
Participants in the AI condition used various general-purpose AI assistants, including ChatGPT, Claude, GitHub Copilot Chat, and Gemini, for purposes beyond test generation. In MarsRover and BowlingScoreKeeper, 60\% and 80\% of participants, respectively, used AI for multiple purposes, including understanding program behavior, planning and generating tests, resolving configuration issues, debugging test failures, and reviewing coverage.
Notably, one participant used AI exclusively for test planning rather than generating complete tests. This participant requested a checklist of testing scenarios and a test-file skeleton with unfinished test cases, explicitly asking AI to leave the implementation for them to complete. Other participants primarily used AI to generate tests, with little separate interaction focused on explanation or troubleshooting. These contrasting behaviors suggest that, even without explicit instructional guidance, students adopted different approaches to AI assistance, particularly in deciding which aspects of testing to delegate to AI and which to perform themselves.

In contrast, participants in the TC condition primarily used TestCompass to examine coverage information and review recommendations while developing or refining their tests, reflecting the tool's more specialized functionality.

\paragraph{Test Execution and Verification Behaviors}
Differences also emerged in how participants tested and verified their testing code. Participants using TestCompass were more likely to execute JUnit tests and inspect their results, while those using general-purpose AI assistants appeared to engage less frequently in these hands-on verification activities.
Specifically, 87.5\% of TC participants performed at least one JUnit execution or result-inspection activity in each task, compared with 50\% of AI participants in MarsRover and 60\% in BowlingScoreKeeper. 
A similar, though smaller, difference was observed in the time spent on these activities. On average, TC participants devoted 9.2\% of classified task time to JUnit execution or result inspection in MarsRover and 10.0\% in BowlingScoreKeeper, compared with 9.5\% and 7.7\%, respectively, for AI participants.
These findings suggest that specialized testing support may help integrate hands-on verification more consistently into students' testing workflows while preserving opportunities to collaborate with AI.

\begin{tcolorbox}[
    colback=white,
    colframe=summaryBorder,
    boxrule=0.5pt,
    arc=2mm,
    left=6pt,
    right=6pt,
    top=5pt,
    bottom=5pt
]
TestCompass participants spent more time directly developing tests and were more likely to execute them and inspect their results, whereas general AI participants devoted more time to interacting with AI assistants across various testing activities. These suggest that feedback-oriented AI support may encourage greater hands-on engagement while still facilitating AI collaboration.
\end{tcolorbox}

\definecolor{qualityComp}{RGB}{242,243,245}     
\definecolor{qualityEff}{RGB}{255,239,225}      
\definecolor{qualityMaint}{RGB}{232,242,234}    
\definecolor{qualityBarrier}{RGB}{230,241,253}  


\begin{table}[t]
\centering
\caption{Test-suite quality by task and condition.}
\label{tab:quality}

\footnotesize
\setlength{\tabcolsep}{2pt}
\renewcommand{\arraystretch}{1.0}

\resizebox{0.85\columnwidth}{!}{%
\begin{tabularx}{1.0\columnwidth}
{@{}>{\raggedright\arraybackslash}X|r|r|r@{}}
\hline
\textbf{Measure}
& \multicolumn{2}{c|}{\textbf{Groups}}
& \textbf{Cliff's $\delta$} \\
\cline{2-3}
& \textit{TC}
& \textit{AI}
& \\
\hline\hline

\multicolumn{4}{c}{\textit{Task 1: MarsRover}} \\
\hline

\rowcolor{qualityComp}
Requirement coverage (\%)
& 85.4 & 90.8 & 0.00 \\

\rowcolor{qualityComp}
Full requirement coverage, n (\%)
& 4 (50\%) & 4 (40\%) & 0.10 \\

\rowcolor{qualityComp}
Line coverage (\%)
& 88.1 & 98.2 & -0.21 \\

\rowcolor{qualityComp}
Branch coverage (\%)
& 72.3
&87.7 & -0.65 \\

\rowcolor{qualityEff}
Mutation coverage (\%)
& 79.2 & 84.9 & -0.25 \\

\rowcolor{qualityMaint}
Maintainability issues
& 2.9 & 4.8 & -0.06 \\

\rowcolor{qualityMaint}
Issues per test method
& 0.36 & 0.19 & 0.43 \\

\rowcolor{qualityMaint}
Test-name quality (1--5)
& 3.1 & 3.7 & -0.45 \\
\rowcolor{qualityMaint}
Redundant test methods
& 0.5 & 1.5 & -0.20\\
\rowcolor{qualityBarrier}
Inventoried test methods
& 12.0
& 28.3 & -0.80 \\

\rowcolor{qualityBarrier}
Erroneous tests (\%)
& 0.0 & 0.3 & -0.10 \\

\hline

\multicolumn{4}{c}
{\textit{Task 2: BowlingScoreKeeper}} \\
\hline

\rowcolor{qualityComp}
Requirement coverage (\%)
& 89.4 & 96.2 & -0.03 \\

\rowcolor{qualityComp}
Full requirement coverage, n (\%)
& 6 (75\%) & 7 (70\%) & 0.05 \\

\rowcolor{qualityComp}
Line coverage (\%)
& 89.4 & 95.7 & 0.13 \\

\rowcolor{qualityComp}
Branch coverage (\%)
& 80.5 & 86.8 & -0.05 \\

\rowcolor{qualityEff}
Mutation coverage (\%)
& 80.8 & 91.5 & -0.18 \\

\rowcolor{qualityMaint}
Maintainability issues
& 6.3 & 5.2 & 0.30 \\

\rowcolor{qualityMaint}
Issues per test method
& \textbf{0.40}$^{*}$
& \textbf{0.19}$^{*}$ & 0.78 \\

\rowcolor{qualityMaint}
Test-name quality (1--5)
& 3.0 & 3.3 & -0.40 \\

\rowcolor{qualityMaint}
Redundant test methods
& 0.6 & 0.5 & -0.11\\

\rowcolor{qualityBarrier}
Inventoried test methods
& \textbf{17.9}$^{*}$
& \textbf{27.6}$^{*}$ & -0.85 \\

\rowcolor{qualityBarrier}
Erroneous tests (\%)
& 11.0 & 2.6 & 0.19 \\

\hline
\end{tabularx}%
}

\par\smallskip
\noindent
\begin{minipage}{\columnwidth}
\scriptsize
\raggedright

{\color{qualityComp}\rule{0.8em}{0.8em}}
Completeness;
{\color{qualityEff}\rule{0.8em}{0.8em}}
Effectiveness;
{\color{qualityMaint}\rule{0.8em}{0.8em}}
Maintainability;
{\color{qualityBarrier}\rule{0.8em}{0.8em}}
Implementation Barriers.

\smallskip
\textit{Note.}
Values are participant-level means, except for full requirement coverage (n, \%).
Issues per test method are normalized by suite size.
Cliff's $\delta$ indicates the rank-based effect size; negative values favor higher AI measurements.
$^{*}$ Indicates a significant between-condition difference (Holm-adjusted $p<0.05$).

\end{minipage}
\end{table}

\subsection{RQ2: How does objective test-suite quality
differ between the conditions?}
\label{sec:rq2}
We evaluated students' test-suite quality across four dimensions: completeness, effectiveness, maintainability, and implementation barriers, as summarized in Table~\ref{tab:quality}.
\paragraph{Completeness}
Across both tasks, participants using general-purpose AI achieved higher mean requirement and code coverage than those using TestCompass, although none of these differences remained statistically significant after multiple-comparison correction.
For MarsRover, AI participants achieved higher mean requirement coverage (90.8\% vs.\ 85.4\%), line coverage (98.2\% vs.\ 88.1\%), and branch coverage (87.7\% vs.\ 72.3\%). 
Despite these differences, four participants in each condition covered all 12 requirements. 
A similar pattern emerged in BowlingScoreKeeper, where AI participants achieved higher mean requirement coverage (96.2\% vs.\ 89.4\%), line coverage (95.7\% vs.\ 89.4\%), and branch coverage (86.8\% vs.\ 80.5\%). 
However, six TestCompass participants and seven AI participants covered all 13 requirements. 
Robustness checks using JaCoCo confirmed that these coverage patterns were consistent across measurement tools and were not substantially influenced by the individual submissions examined.

Notably, both conditions achieved similar rates of complete requirement coverage, despite TestCompass providing recommendations rather than generating executable tests. These findings suggest
that while direct test generation may help novices
achieve broader code coverage, it is not necessarily required to achieve complete requirement coverage.
Recommendation-based support may be sufficient to
guide novices toward relevant testing behaviors,
even when they are responsible for implementing
the tests themselves.

\paragraph{Effectiveness}
We assessed test effectiveness using mutation coverage, which measures a test suite's ability to detect injected faults. 
General AI participants achieved higher mean mutation coverage than TestCompass participants on both MarsRover (84.9\% vs.\ 79.2\%) and BowlingScoreKeeper (91.5\% vs.\ 80.8\%), although neither difference was statistically significant. For one TestCompass participant, two tests with compilation errors were commented out to enable mutation analysis. Assigning this participant a mutation score of zero did not change the statistical conclusions. Thus, although participants using TestCompass authored all tests themselves, guided only by its recommendations, we found no statistically detectable disadvantage in fault detection.


\paragraph{Maintainability}
We assessed test-suite maintainability through test-name quality and the number of maintainability issues. Across both tasks, AI participants achieved higher mean test-name quality ratings than TC participants, although neither difference was statistically significant.
The number of maintainability issues showed mixed results. 
AI participants produced more issues in MarsRover (4.8 vs.\ 2.9), whereas TC participants produced more in BowlingScoreKeeper (6.3 vs.\ 5.2). However, because AI participants generally developed larger test suites, we normalized issue counts by the number of test methods. 
After normalization, AI suites exhibited fewer issues per test method in both MarsRover (0.19 vs.\ 0.36) and BowlingScoreKeeper (0.19 vs.\ 0.40), although only the latter difference was statistically significant ($p=0.044$).
Regarding test redundancy, TC participants produced fewer redundant test methods on average in MarsRover (0.5 vs.\ 1.5), while both conditions showed similar levels of redundancy in BowlingScoreKeeper (0.6 vs.\ 0.5). Neither difference was statistically significant.

Overall, general-purpose AI assistance was associated with higher test-name quality ratings and lower maintainability issue density, although most differences were not statistically significant. 
These findings suggest that feedback-oriented support could be extended beyond identifying weaknesses in students' test suites to helping them understand and apply principles of effective test design, including clear test naming and maintainable test structure.

\paragraph{Implementation Barriers}
We examined implementation-related challenges through test-suite size and the proportion of erroneous tests. Notably, although TestCompass participants achieved similar rates of complete requirement coverage, they did so with fewer test methods. In MarsRover, TC participants submitted an average of 12.0 test methods, compared with 28.3 for AI participants. A similar pattern emerged in BowlingScoreKeeper (17.9 vs.\ 27.6), where the difference was statistically significant ($p=0.012$).
We also found no statistically significant differences in test correctness. In MarsRover, the mean participant-level confirmed-error rate was 0.0\% for TC participants and 0.3\% for AI participants. In BowlingScoreKeeper, the corresponding rates were 11.0\% and 2.6\%, respectively.

Overall, these findings suggest that TestCompass's recommendation-based support can help novices achieve broad coverage with fewer, independently authored tests, potentially improving testing efficiency. Although general AI participants produced larger test suites, these additional tests did not yield statistically significant improvements in coverage or fault detection. This highlights the potential of feedback-oriented support to guide efficient test development while keeping novices actively engaged in designing and implementing their own tests.
Table~\ref{tab:quality}
summarizes the testing quality results for both tasks.
\begin{tcolorbox}[
    colback=white,
    colframe=summaryBorder,
    boxrule=0.5pt,
    arc=2mm,
    left=6pt,
    right=6pt,
    top=5pt,
    bottom=5pt
]
TestCompass participants achieved similar requirement coverage with significantly fewer test methods than general AI participants. These findings suggest that feedback-based AI support can guide novices toward comprehensive and effective testing while keeping them actively engaged in implementing their own tests.
\end{tcolorbox}

\subsection{RQ3: What are students' subjective
experiences with the tools?}
\label{sec:rq3}

\paragraph{Self-Reported Tool Use}
Survey responses revealed differences in how participants incorporated AI support into their testing workflows. 
While identifying what to test next was common in both conditions (TC: 75\%; AI: 70\%), TC participants more frequently sought guidance on whether their testing was sufficient (62.5\% vs.\ 20\%). Six TC participants (75\%) also reported consulting TestCompass before submission, suggesting that they used it to review testing completeness.
In contrast, AI participants reported more frequent tool consultation throughout testing (80\% vs.\ 25\%), suggesting a more continuous role for general-purpose AI in their workflows. These findings suggest that different forms of AI support may play distinct roles in students' testing workflows. While general-purpose AI provides assistance throughout test development, feedback-oriented support may encourage students to periodically evaluate their progress, identify remaining testing gaps, and reflect on whether their tests are sufficient.

\paragraph{Perceived Helpfulness and Recommendation}
Participants perceived general-purpose AI as more helpful than TestCompass, although higher perceived helpfulness did not necessarily translate into stronger recommendations. All AI respondents rated their tool's helpfulness as 4 or 5 on a five-point scale, compared with three of eight TC respondents (100\% vs.\ 37.5\%; $p=0.007$). However, ratings of whether participants would recommend the tool to future students were more similar, with 60\% of AI respondents and 50\% of TC respondents giving ratings of at least 4. Notably, three AI participants (30\%) gave low recommendation ratings (1 or 2) despite rating the tool's helpfulness highly.

These findings highlight a distinction between perceived helpfulness and willingness to recommend a tool. Moreover, the higher helpfulness ratings for general-purpose AI were not accompanied by statistically significant differences in most objective test-suite quality measures (Sec.~\ref{sec:rq2}). This suggests that perceived helpfulness alone may not fully reflect the effectiveness of AI support in programming education.

\paragraph{Self-Assessed Knowledge and Confidence}
Despite differences in perceived tool helpfulness, participants in both conditions reported little change in their unit-testing knowledge or confidence after completing the tasks. Mean self-assessed knowledge ratings remained at 3.0 before and after the session in both conditions. Confidence ratings were similarly stable, with a slight increase among TC participants.
These findings suggest that, despite receiving AI assistance during testing, participants did not report noticeable short-term improvements in their knowledge or confidence within a single session. Table~\ref{tab:survey} summarizes the survey responses.

\paragraph{Reported Challenges}
Across both conditions, participants reported common difficulties identifying additional test cases, understanding task specifications, determining expected outputs, and deciding when testing was sufficient.
TC participants also faced challenges in interpreting and applying TestCompass’s recommendations, and expressed a need for guidance that explains the reasoning behind effective test design. For example, one participant requested:

\quoteline \textit{``A clearer understanding of what good unit testing looks like vs bad.''}

\noindent Another wanted guidance on what to consider when designing tests rather than simply receiving instructions:

\quoteline \textit{``some further guidance \ldots\ as to what to think and expect when writing unit tests.''}

\noindent These responses suggest that identifying testing gaps alone may not be sufficient for novices, who may also need explanations to understand and translate recommendations into effective tests.

In contrast, AI participants valued the convenience of generating tests and resolving formatting issues, although some encountered difficulties with project setup or generated tests that failed during execution. Notably, one participant expressed an interest in learning to program independently:

\quoteline \textit{``How to code without using AI systems?''}

\noindent These responses highlight a potential tension between providing immediate assistance and helping novices develop independent testing judgment. While general-purpose AI can simplify test development, feedback-based support leaves more of the testing decisions and implementation to students. However, participants' feedback suggests that such support may benefit from clearer, more actionable explanations that help novices understand not only what to test, but also how and why.

\begin{tcolorbox}[
    colback=white,
    colframe=summaryBorder,
    boxrule=0.5pt,
    arc=2mm,
    left=6pt,
    right=6pt,
    top=5pt,
    bottom=5pt
]
Participants found general AI more helpful and consulted it more
frequently during test development, whereas TestCompass was used mainly to guide
testing decisions and assess completeness. 
Neither tool improved participants’ perceived knowledge or confidence, suggesting that neither form of AI assistance necessarily leads to short-term gains in perceived learning.
\end{tcolorbox}

\begin{table}[t]
\centering
\caption{Participant survey ratings by tool condition.}
\label{tab:survey}

\footnotesize
\setlength{\tabcolsep}{6pt}
\renewcommand{\arraystretch}{1.08}

\begin{tabular}{@{}lrr@{}}
\toprule
& \multicolumn{2}{c}{\textbf{Groups}} \\
\cmidrule(lr){2-3}
\textbf{Measure}
& \textit{TC}
& \textit{AI} \\
\midrule

Tool helpfulness
& \textbf{3.3}$^{*}$
& \textbf{4.7}$^{*}$ \\

Willingness to recommend
& 3.4 & 3.6 \\
\hline
\addlinespace[1pt]
Testing knowledge (pre)
& 3.0 & 3.0 \\

Testing knowledge (post)
& 3.0 & 3.0 \\
\hline
\addlinespace[1pt]
Testing confidence (pre)
& 2.8 & 2.9 \\

Testing confidence (post)
& 2.9 & 2.9 \\

\bottomrule
\end{tabular}

\par\smallskip
\begin{minipage}{0.95\columnwidth}
\scriptsize
\raggedright
\textit{Note.}
All measures use a 1--5 scale. Knowledge and confidence are self-reported unit-testing ratings, collected before (pre) and after (post) the study.
$^{*}$ Indicates a significant between-condition difference
(Holm-adjusted $p<0.05$).
\end{minipage}
\end{table}

\section{Threats to Validity}
\label{sec:limitations}

\subsection{Internal Validity}
Participants were randomly assigned to the TC or AI condition, and their self-reported prior programming and testing experience was comparable across conditions. To reduce variation in educational background, participants were primarily recruited from the same course, with one exception from a senior-level course. Nevertheless, unmeasured individual differences in programming proficiency, testing skills, or familiarity with AI tools may have influenced their performance.

The general-purpose AI condition was not fully standardized, as participants were allowed to use different AI assistants due to IRB restrictions. Although initially designated as the ChatGPT condition, participants also used tools such as Claude, Gemini, and GitHub Copilot Chat. Differences in model capabilities, response quality, and interaction patterns may therefore have introduced variation. To account for this heterogeneity in our interpretation, we frame the study as a comparison between two forms of AI support, namely general-purpose AI assistants capable of generating executable tests and a feedback-oriented tutor that provides testing recommendations. Nevertheless, differences among general-purpose AI assistants remain a potential source of variability.

Additionally, both tasks were completed in a fixed order, making it difficult to distinguish task-specific effects from potential practice, fatigue, or carryover effects. Future studies could counterbalance task order and incorporate breaks between tasks to mitigate these concerns.
Finally, screen recording may have influenced participants' behavior by increasing performance pressure, although the same requirement applied to both conditions. To minimize this effect, we informed participants that they should do their best and that their performance would not affect their course grades.

\subsection{External Validity}
Most participants were recruited from a single sophomore-level computer science course at one institution, with prior exposure to Java, JUnit, and introductory unit testing. The relatively small sample size, partly due to recruitment challenges, may further limit the generalizability of our findings to students from other institutions, with different educational backgrounds or experience levels, and to professional developers.
Nevertheless, the design of TestCompass primarily targets novice programmers learning unit testing. 
To strengthen our findings despite the limited sample, we examined testing behaviors and outcomes across two tasks using multiple complementary measures. We also employed statistical tests appropriate for small samples and reported effect sizes to support interpretation. 
Second, our study focused on two Java unit-testing tasks completed in IntelliJ IDEA. Other testing domains, programming languages, development environments, or testing frameworks may introduce different challenges and interaction patterns. Thus, the findings and the current implementation of TestCompass may not generalize directly to other testing contexts.
Finally, the study lasted approximately one and a half hours, capturing short-term testing behaviors, performance, and perceptions rather than long-term learning, retention, or transfer of testing skills. Future longitudinal studies are needed to examine whether different forms of AI support influence students' testing practices and skill development over time.

\subsection{Construct Validity}
We evaluated test-suite quality across four dimensions: completeness, effectiveness, maintainability, and implementation barriers. Although these dimensions capture important aspects of testing performance, they do not fully represent the underlying constructs. Requirement, line, and branch coverage measure the extent of testing but do not necessarily reflect test correctness or semantic quality. Mutation coverage approximates fault-detection effectiveness but depends on the generated mutants and mutation operators. Excluding tests with compilation errors may affect measured mutation coverage, although assigning the affected suite a zero score did not change our statistical conclusions.
Maintainability was assessed using static-analysis issues, test-name quality, and test redundancy, which may not fully capture long-term maintainability. Similarly, the number of submitted tests and confirmed errors provides only a partial view of implementation challenges.

Manual coding and qualitative analysis also involved researcher judgment. To reduce subjectivity, we applied consistent coding criteria and involved a second researcher in reviewing annotations and themes, although some interpretive bias may remain.
Survey measures, including self-assessed knowledge, confidence, helpfulness, and willingness to recommend the tool, may be subject to response bias and do not directly measure learning gains. Finally, interaction analysis relied on screen recordings and observable behaviors, which cannot fully capture participants' underlying reasoning or decision-making processes.

\section{Discussion}
\label{sec:discussion}

\subsection{Teaching Software Testing in the AI Era}

As generative AI increasingly produces both code and its tests, testing education must place greater emphasis on students’ ability to evaluate whether tests adequately cover intended behaviors and interpret what their results reveal.
Our findings highlight the potential value of different forms of AI assistance. Participants using general-purpose AI produced substantially larger test suites and generally achieved higher mean coverage. However, comparable proportions of TestCompass participants covered all task requirements despite receiving only recommendations rather than executable tests, and no coverage difference remained significant after multiple-comparison correction. Interaction patterns also differed: TestCompass participants spent more of their time writing, revising, and inspecting tests, whereas general AI participants spent more time consulting their assistants. Although these differences do not establish better learning, they suggest that the form of AI support shapes how students engage with testing.

These findings suggest that educators could select AI tools based on their instructional goals rather than relying on a single approach. General AI may be useful for exploring unfamiliar concepts and examples, while recommendation-based AI tutors may be better suited to hands-on exercises that require students to implement and evaluate tests themselves. This approach may help students develop the skills to work effectively with AI tools while remaining actively engaged in the testing process rather than passively relying on AI-generated solutions.
Regardless of the tool, educators should encourage students to justify their test choices, explain expected outcomes, and identify weaknesses in AI-generated tests. Such activities may help students develop independent testing judgment while benefiting from AI assistance.

\subsection{Designing AI Tutors That Bridge Guidance
and Implementation}

Our findings highlight opportunities for improving
testing-specific AI support. TestCompass was designed
to identify testing gaps without directly generating
tests, allowing students to remain responsible for
test implementation. However, participants sometimes
struggled to interpret lengthy recommendations,
translate them into executable tests, and resolve
execution problems. In contrast, general-purpose AI
participants produced substantially more tests and
rated their tools as more helpful. These findings
highlight a challenge for recommendation-based
support: knowing what to test does not necessarily
mean knowing how to implement it.

Rather than choosing between providing recommendations
and generating complete tests, future AI tutors could
offer different levels of assistance based on students'
needs. Building on prior work on conceptual feedback
and progressive testing hints~\cite{cordova2021comparison,
perretta2025instructor}, a tutor could initially
help students identify relevant requirements and
reason about expected outcomes. When students
encounter difficulties, it could offer more concrete
implementation guidance or help diagnose test failures,
such as distinguishing incorrect assertions from
implementation defects. This progressive approach
could bridge the gap between understanding testing
recommendations and putting them into practice,
while preserving opportunities for students to
develop and verify tests themselves.

\subsection{Feedback Must Support Oracle and Domain Reasoning}

An important challenge emerging from our findings is that knowing what to test does not mean knowing the correct result. In both conditions, participants struggled to understand task specifications, determine expected outputs, and judge whether their tests were sufficient. This was especially true for BowlingScoreKeeper, where testing strikes, spares, and final-frame behavior requires understanding how rolls contribute to the final score. Although general-purpose AI provides both test code and explanations, students may use the generated tests without examining the reasoning behind their expected outcomes.
This distinction matters because identifying a relevant test case does not guarantee a correct assertion. Students may recognize which behavior to test yet misunderstand the specification or compute an incorrect expected result. Similarly, an AI-generated test may run successfully without checking the correct behavior. TestCompass addresses this by including validated expected outcomes in its recommendations. However, participants’ difficulties suggest that providing the expected result alone may not help novices understand why it is correct.

Future AI testing tutors should therefore help students reason about expected outcomes rather than simply provide them. For example, feedback could explain which requirement a test addresses and how its expected result is derived. Given participants’ difficulty with lengthy recommendations, such explanations should be concise and revealed progressively as needed. This could help students understand the reasoning behind their own tests and critically evaluate AI-generated tests instead of accepting their expected outputs unverified.

\subsection{Perceived Helpfulness Does Not Necessarily
Reflect Learning Support}

Participants rated general-purpose AI as more helpful
than TestCompass, yet this difference was not accompanied
by a statistically significant advantage in requirement
coverage. Neither condition showed consistent improvements
in self-assessed testing knowledge or confidence.
Willingness to recommend the tools also did not differ
significantly, and some AI participants gave low
recommendation ratings despite finding the tool helpful.
One participant even expressed a desire to learn
programming without relying on AI. These findings
suggest that perceived helpfulness may reflect the
convenience of completing an immediate task rather
than the extent to which a tool supports learning.

This distinction has implications for evaluating
educational AI tools. While general-purpose AI can
provide immediate solutions, recommendation-based
tutors such as TestCompass require students to
interpret guidance and implement tests themselves.
However, our single-session study cannot determine
whether this additional involvement leads to better
learning outcomes. Future evaluations should therefore
look beyond task performance and perceived helpfulness
to examine whether students can independently design
tests, justify expected outcomes, and identify errors
in AI-generated tests after assistance is removed.

\section{Conclusion}
\label{sec:conclusion}

As generative AI increasingly supports both code and test generation, testing education needs approaches that provide meaningful assistance while keeping students actively engaged in designing and verifying their own tests.
In this study, we introduced TestCompass, a feedback-oriented AI tutor that provides targeted testing recommendations without generating executable tests, and compared it with general-purpose AI assistance in a study involving 18 novice programmers.
We found that participants using general-purpose AI produced larger test suites, generally achieved higher coverage, and perceived their tools as more helpful. However, most differences in objective test-suite quality were not statistically significant, and both conditions achieved similar rates of complete requirement coverage. TestCompass participants spent more time developing tests and were more likely to execute and inspect their results. Neither condition showed consistent improvements in self-assessed knowledge or confidence.
These findings suggest that feedback-oriented AI can support meaningful testing outcomes while preserving opportunities for students to develop and verify their own tests. However, participants' difficulties in interpreting recommendations and determining expected outcomes highlight the need for additional implementation guidance. Future work should explore adaptive AI tutors that integrate testing recommendations with more implementation and verification support, while examining their effects on students' independent testing skills over time. Ultimately, TestCompass demonstrates the potential of AI to support effective testing practice without taking test development away from students.


\newpage
\bibliographystyle{IEEEtran}
\bibliography{my}
\end{document}